# HARMONIC ANALYSIS ON DIRECTED NETWORKS VIA A BIORTHOGONAL LAPLACIAN CALCULUS FOR NON-NORMAL DIGRAPHS

**Chandrasekhar Gokavarapu** *Lecturer in Mathematics, Government College (Autonomous), Rajahmundry, A.P., India :: :* chandrasekhargokavarapu@gmail.com
**Dr Komala Lakshmi Chinnam** *Lecturer in Physics, Government College (Autonomous), Rajahmundry, A.P., India*

## Abstract

Spectral graph signal processing is traditionally built on self-adjoint Laplacians, where orthogonal eigenbases yield an energy-preserving Fourier transform and a variational frequency ordering via a real Dirichlet form. Directed networks break self-adjointness: the combinatorial directed Laplacian $L = D_{out} - A$ is generally non-normal, so eigenvectors are non-orthogonal and classical Parseval identities and Rayleigh-quotient orderings do not apply. This paper develops a Laplacian-centric harmonic analysis for directed graphs that remains exact at the algebraic level while explicitly quantifying the geometric distortion induced by non-normality. We (i) define a Biorthogonal Graph Fourier Transform (BGFT) for $L$ using dual left/right eigenbases and show that vertex energy equals a Gram-metric quadratic form in BGFT coordinates,

(ii) introduce a directed variational semi-norm $TV_G(x) = \|Lx\|_2^2$ and prove sharp two-sided BGFT-domain bounds controlled by singular values of the eigenvector matrix, and (iii) derive sampling and reconstruction guarantees with explicit stability constants that separate sampling-set informativeness from eigenvector geometry. Finally, we provide reproducible simulations comparing a normal directed cycle to perturbed non-normal digraphs and show that filtering and reconstruction robustness track $\kappa(V)$ and the Henrici departure-from-normality $\Delta(L)$, validating the theoretical predictions.

*Keywords:* Directed graph signal processing, Combinatorial directed Laplacian, Biorthogonal graph Fourier transform, Non-normal matrix analysis, Sampling and reconstruction

*2000 MSC:* 05C50, 15A18, 94A12, 65F15

## 1. Introduction

Spectral graph theory is a standard foundation for graph signal processing (GSP), where a graph Fourier transform (GFT) diagonalizes a chosen graph operator and enables filtering, denoising, and sampling on networks [1, 2, 3]. In the undirected case, symmetry yields a self-adjoint Laplacian with a complete orthonormal eigenbasis; the GFT is an isometry and energy, smoothness, and frequency ordering are unified through the Dirichlet quadratic form.

Directed networks are different: the combinatorial directed Laplacian $L = D_{out} - A$ is typically *non-normal* ($LL^* \neq L^*L$). As a result, eigenvectors are non-orthogonal, the spectral representation is not an isometry, and small perturbations in spectral coefficients can cause large reconstruction errors. This is not a technical nuisance but a structural phenomenon governed by eigenvector geometry and pseudospectral behavior [10, 11].

*Positioning and novelty.*. Existing directed GSP literature often (a) symmetrizes the problem, losing directionality [4], or (b) defines frequency via adjacency/Jordan calculus [5, 6], with additional alternatives including magnetic Laplacians and optimization-based operators [9, 7, 8]. This paper contributes a *Laplacian-variational* directed harmonic analysis that is:

- *Algebraically exact*: analysis/synthesis is exact under diagonalizability, and diagonal filtering is well-defined in BGFT coordinates;

- *Geometrically quantified*: all energy and smoothness identities are stated with explicit Gram-metric and conditioning constants that measure nonnormality-induced distortion;

- *Operational*: sampling and reconstruction statements are given with stability constants that separate sampling-set informativeness from eigenvector non-orthogonality.

Compared with our earlier adjacency-based formulation [12], the present work

(i) uses $L$ to ground frequency in directed variation, (ii) introduces sharp two-sided BGFT bounds for $\|Lx\|_2$ with explicit conditioning constants, and (iii) develops sampling/reconstruction bounds tailored to oblique Laplacian spectral subspaces.

*1.1. Contributions*

1. **Biorthogonal Laplacian GFT and Gram-metric Parseval law.** We construct the BGFT for the directed Laplacian and prove that vertex energy equals a Gram-metric quadratic form in BGFT coordinates; this yields exact energy bounds in terms of singular values and the condition number $\kappa(V)$.
2. **Directed variation and sharp BGFT-domain smoothness bounds.** We define directed smoothness by $TV_G(x) = \|Lx\|_2^2$ and prove two-sided inequalities that become equalities in the normal case, quantifying when eigenvalue magnitudes behave as directed frequencies.
3. **Sampling/reconstruction with explicit stability constants.** We generalize bandlimited sampling to oblique Laplacian spectral subspaces and give exact recovery and noise sensitivity bounds that separate the roles of $(P_M V_\Omega)^\dagger$ and $V_\Omega$.
4. **Reproducible experiments and quantitative non-normality metrics.** We provide simulations that compute spectra, conditioning, Henrici departure-from-normality, and reconstruction errors on controlled digraph families, demonstrating consistency with theory.

*1.2. Organization*

Section 2 fixes conventions and non-normality indices. Section 3 defines the BGFT for $L$ and proves energy identities. Section 4 develops directed variation and frequency ordering. Section 5 states sampling and reconstruction results. Section 6 discusses stability mechanisms and practical computation. Section 7 presents experiments, followed by conclusions.

## 2. Preliminaries

### *2.1. Directed graphs, adjacency, and out-degree*

Let $G = (V,E,w)$ be a directed weighted graph, $|V| = n$, with adjacency $A \in \mathbb{R}^{n\times n}$ given by

$$A_{ij} = w(i,j) \quad \text{if } (i,j) \in E, \qquad A_{ij} = 0 \text{ otherwise.}$$

Thus edges are oriented $i \to j$ and the out-degrees are

$$d_i^{\text{out}} = \sum_{j=1}^{n} A_{ij}, \qquad D_{\text{out}} = \operatorname{diag}(d_1^{\text{out}},\ldots,d_n^{\text{out}}).$$

### *2.2. Combinatorial directed Laplacian*

**Definition 2.1** (Directed Laplacian)**.** The combinatorial directed Laplacian is $L := D_{\text{out}} - A$.

**Proposition 2.2** (Row-sum zero)**.** *For any directed graph (with any weights),* $L\mathbf{1} = 0$.

*Proof.* The $i$th entry of $L\mathbf{1}$ equals $d^{\text{out}}_i - \sum_j A_{ij} = 0$ by definition.

**Remark 2.3** (Non-self-adjointness)**.** If $A \neq A^T$, then typically $L \neq L^T$ and $L$ is not self-adjoint. Orthogonality of eigenvectors and Rayleigh-quotient variational orderings may fail; this motivates a biorthogonal calculus.

### *2.3. Asymmetry and non-normality indices*

**Definition 2.4** (Asymmetry index)**.** For any $M$, define $\alpha(M) := \|M - M^\top\|_F / \|M\|_F$ (with $\alpha(0) = 0$).

**Definition 2.5** (Commutator-based departure from normality)**.** For any $M$, define $\delta(M) := \|MM^* - M^*M\|_F / \|M\|_F^2$ (with $\delta(0) = 0$).

**Definition 2.6** (Henrici departure from normality)**.** For any $M \in \mathbb{C}^{n\times n}$ with eigenvalues $\{\lambda_k\}_{k=1}^n$,

$$\Delta(M) := \sqrt{\|M\|_F^2 - \sum_{k=1}^{n} |\lambda_k|^2}.$$

Normal matrices satisfy $\Delta(M) = 0$ [10, 11].

## 3. Biorthogonal Graph Fourier Transform for the directed Laplacian

### *3.1. Biorthogonal spectral decomposition*

We assume $L$ is diagonalizable,[1] so

$$L = V\Lambda V^{-1},$$

where $V = [v_1,\ldots,v_n]$ contains right eigenvectors ($Lv_k = \lambda_k v_k$) and $\Lambda = \operatorname{diag}(\lambda_1,\ldots,\lambda_n)$.

[1] Defective cases can be handled with Schur or Jordan calculus; the resulting nonorthogonality effects are typically stronger and are naturally studied through pseudospectra [10].

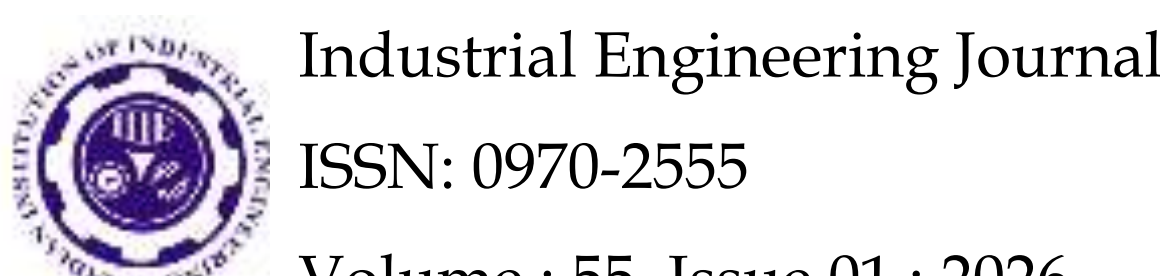

Define the dual (left) basis via

$$U := (V^{-1})^* \quad \Longleftrightarrow \quad U^* = V^{-1}.$$

Then the columns $u_k$ of $U$ satisfy $L^* u_k = \overline{\lambda_k} u_k$ and biorthogonality holds:

$$u_j^* v_i = \delta_{ij}.$$

*3.2. Transform pair*

**Definition 3.1** (BGFT)**.** For a graph signal $x \in \mathbb{C}^n$, its BGFT coefficients are

$$\hat{x} = U^* x = V^{-1} x, \qquad \hat{x}_k = u_k^* x.$$

**Definition 3.2** (Inverse BGFT)**.** The inverse BGFT is $x = V\hat{x} = \sum_{k=1}^{n} \hat{x}_k v_k$.

*3.3. Gram-metric Parseval identity and energy bounds*

Non-orthogonality induces a metric distortion in the spectral domain. Let $M := V^*V$ be the Gram matrix of the right eigenvectors.

**Theorem 3.3** (Exact energy identity)**.** *For any $x \in \mathbb{C}^n$ with BGFT coefficients $\hat{x} = V^{-1}x$,*

$$\|x\|_2^2 = \hat{x}^* M \hat{x}.$$

*Proof.* Since $x = V\hat{x}$, we have $\|x\|_2^2 = \langle V\hat{x}, V\hat{x}\rangle = \hat{x}^*(V^*V)\hat{x}$. □

**Corollary 3.4** (Two-sided Parseval bounds)**.** Let $\sigma_{\min}(V), \sigma_{\max}(V)$ denote the smallest and largest singular values of $V$. Then

$$\sigma_{\min}^2(V)\,\|\hat{x}\|_2^2 \le \|x\|_2^2 \le \sigma_{\max}^2(V)\,\|\hat{x}\|_2^2.$$

Equivalently, energy distortion is controlled by $\kappa(V) = \sigma_{\max}(V)/\sigma_{\min}(V)$.

*3.4. DC component and mean mode*

By Proposition 2.2, $\lambda = 0$ is always an eigenvalue with right eigenvector $\mathbf{1}$. Thus the Laplacian isolates a natural "DC" mode (constant signal), without requiring regularity assumptions that appear in adjacency-based formulations.

## 4. Directed variation and frequency ordering

*4.1. Directed smoothness semi-norm*

The quadratic form $x^*Lx$ is generally complex for non-self-adjoint $L$. Instead, we measure directed variation by the magnitude of the Laplacian response.

**Definition 4.1** (Directed total variation)**.**

$$TV_G(x) := \|Lx\|_2^2 = x^* L^* L x.$$

*4.2. BGFT-domain bounds for variation*

In undirected GSP, $TV(x)$ equals a weighted sum of $|\lambda_k|^2|\hat{x}_k|^2$. The directed, non-normal case inherits this relation up to sharp conditioning constants.

**Theorem 4.2** (Sharp two-sided BGFT variation bounds)**.** *Let* $x = V\hat{x}$ *and* $L = V\Lambda V^{-1}$*. Then*

$$\sigma_{\min}^2(V)\sum_{k=1}^{n}|\lambda_k|^2|\hat{x}_k|^2 \;\le\; \|Lx\|_2^2 \;\le\; \sigma_{\max}^2(V)\sum_{k=1}^{n}|\lambda_k|^2|\hat{x}_k|^2.$$

*Proof.* We have $Lx = V\Lambda\hat{x}$. For any vector $z$, $\sigma_{\min}(V)\|z\|_2 \le \|Vz\|_2 \le \sigma_{\max}(V)\|z\|_2$. Let $z = \Lambda\hat{x}$ and square the resulting inequalities. □

**Corollary 4.3** (Frequency ordering and tightness)**.** Ordering modes by nondecreasing $|\lambda_k|$ minimizes the upper bound in Theorem 4.2. The interpretation of $|\lambda_k|$ as a directed "frequency" is tight when $\kappa(V)$ is moderate and becomes loose in strongly non-normal regimes.

## 5. Sampling and reconstruction for $L$-bandlimited signals

*5.1. Bandlimited model*

Let $\Omega \subset \{1,...,n\}$ be an index set of size $K$ representing low directed frequencies (small $|\lambda_k|$). Define

$$V_\Omega := [v_k]_{k\in\Omega} \in \mathbb{C}^{n\times K}, \qquad B_\Omega := \operatorname{span}(V_\Omega).$$

A signal is $\Omega$-bandlimited if $x \in B_\Omega$.

*5.2. Exact recovery and stability*

Let $M \subseteq \{1,...,n\}$ be a sampling set of vertices, and let $P_M \in \{0,1\}^{m\times n}$ be the restriction operator extracting entries indexed by $M$.

**Theorem 5.1** (Exact recovery)**.** *If* $x = V_\Omega c \in B_\Omega$ *and* $B := P_M V_\Omega$ *has full column rank K, then x is uniquely determined by* $y = P_M x$ *and can be recovered by*

$$\hat{c} = B^{\dagger}y, \qquad \hat{x} = V_\Omega\hat{c}.$$

**Definition 5.2** (Sampling stability constant)**.** Assuming $\operatorname{rank}(B) = K$, define $\gamma(M,\Omega) := \sigma_{\min}(B) > 0$.

**Theorem 5.3** (Noise sensitivity)**.** *If* $y = P_M x + \eta$ *with noise* $\eta \in \mathbb{C}^m$ *and* $\hat{x}$ *is reconstructed by least squares as in Theorem 5.1, then*

$$\|\hat{x} - x\|_2 \le \|V_\Omega\|_2 \frac{\|\eta\|_2}{\gamma(M,\Omega)}.$$

**Remark 5.4** (Separation of instability mechanisms)**.** The bound separates

(i) *sampling geometry* via $\gamma(M,\Omega)^{-1}$ and (ii) *eigenvector geometry* via $\|V_\Omega\|_2$ (non-orthogonality/scaling). This separation is specific to the directed, oblique subspace setting.

## 6. Stability, non-normality, and practical computation

### *6.1. Reconstruction stability under spectral perturbations*

**Theorem 6.1** (Coefficient-to-signal amplification)**.** *Let* $\hat{x}$ *be BGFT coefficients of* $x = V\hat{x}$*. If coefficients are perturbed to* $\hat{x} + \eta$*, then*

$$\frac{\|V(\hat{x}+\eta) - V\hat{x}\|_2}{\|x\|_2} \leq \kappa(V)\,\frac{\|\eta\|_2}{\|\hat{x}\|_2}.$$

*Proof.* The error is $V\eta$, so $\|V\eta\|_2 \leq \|V\|_2\|\eta\|_2$. Also $\|\hat{x}\|_2 = \|V^{-1}x\|_2 \leq \|V^{-1}\|_2\|x\|_2$. Combine and rearrange. □

### *6.2. Stable computation of BGFT (recommended)*

Computing $V^{-1}$ explicitly can be unstable when $\kappa(V)$ is large. A standard remedy is to use a numerically stable factorization and avoid forming $V^{-1}$.

**Remark 6.2** (Non-normality as a design constraint)**.** In directed filtering and sampling tasks, $\kappa(V)$ and $\Delta(L)$ behave as *intrinsic difficulty indices*. Large values imply that stable spectral filtering may require (i) regularized filter design, (ii) Schur-based spectral methods, or (iii) operator choices other than $L$ for the application at hand.

**Algorithm 1** Numerically stable BGFT computation (outline)

**Require:** Directed Laplacian $L \in \mathbb{R}^{n\times n}$, signal $x \in \mathbb{C}^n$

**Ensure:** BGFT coefficients $\hat{x}$

1: Optionally scale/balance $L$ to reduce non-normal effects [11] 2: Compute eigen-decomposition $L = V\Lambda V^{-1}$ (or Schur form if needed)

3: Solve the linear system $V\hat{x} = x$ for $\hat{x}$ (do *not* form $V^{-1}$)

4: **return** $\hat{x}$

Table 1: Non-normality and conditioning metrics for the instances used in Figure 1 (computed by make_figures.py with seed 20251221).

| Graph | $\kappa(V)$ | $\Delta(L)$ | $\alpha(L)$ | $\delta(L)$ |
|---|---|---|---|---|
| Directed cycle | 1 | 0 | 1 | 0 |
| Perturbed cycle | 16.80157684 | 5.981556651 | 0.4910602974 | 0.06508077683 |

## 7. Experimental validation

### *7.1. Setup and reproducibility*

We compare two digraph families with $n = 20$ nodes:

1. **Directed cycle** (unweighted): $1 \to 2 \to \cdots \to n \to 1$. This $L$ is non-symmetric but normal, yielding an orthogonal eigenbasis.

2. **Perturbed cycle**: starting from the directed cycle, add random directed edges independently with probability $p = 0.2$ and weight $w = 0.8$, increasing non-normality.

All plots in this paper are generated by the included script (make_figures.py) with a fixed random seed (see repository note in Data Availability).

*7.2. Spectra and non-normality metrics*

Figure 1 shows eigenvalues of $L$ in the complex plane for a representative instance of each family.

Table 1 reports the computed non-normality metrics for the same instances (as produced by the script).

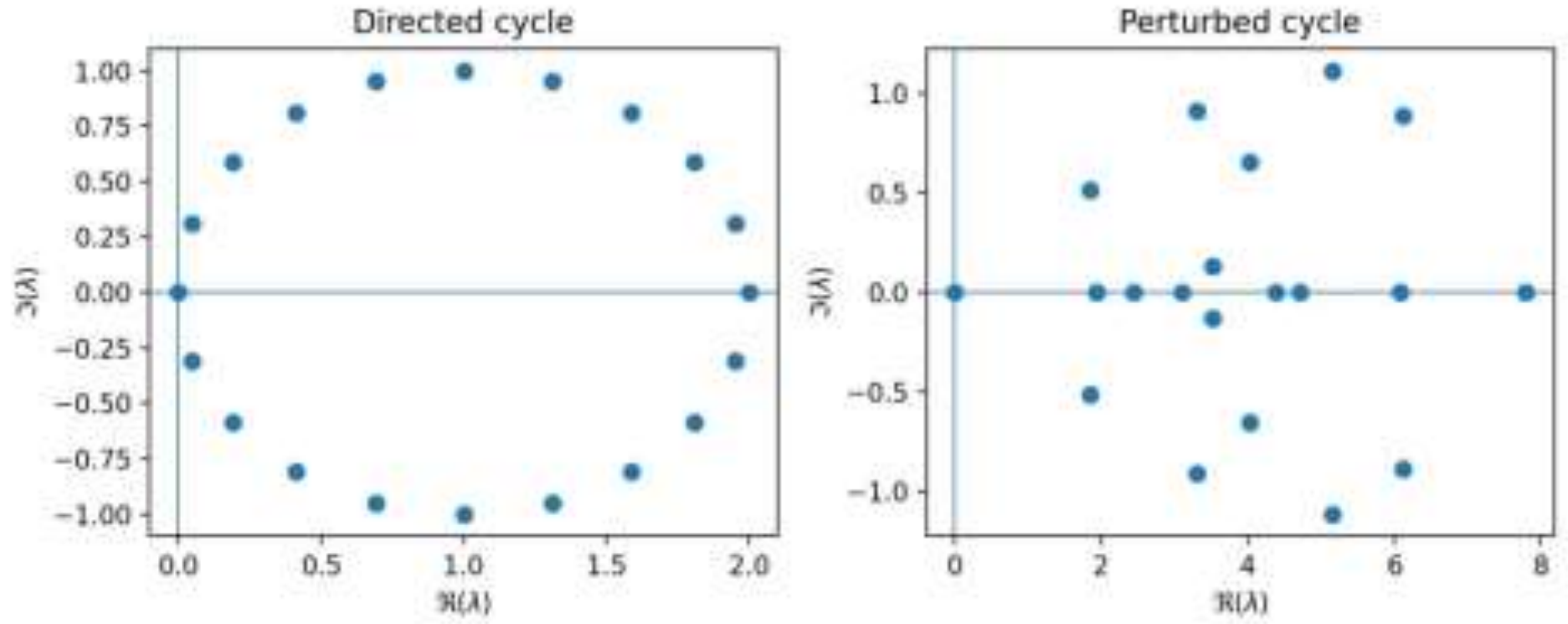

Figure 1: Spectra of the directed Laplacian for a directed cycle (left) and a perturbed cycle (right). Non-normal perturbations visibly deform spectral geometry and typically increase $\kappa(V)$ and $\Delta(L)$.

*7.3. Filtering and reconstruction stability*

We generate a $K$-bandlimited signal using the $K = 5$ lowest-$|\lambda|$ modes, add complex Gaussian noise, and reconstruct via ideal low-pass filtering in BGFT coordinates. Figure 2 shows reconstruction error versus input noise level. The observed gap between curves is consistent with Theorem 6.1 and grows with $\kappa(V)$.

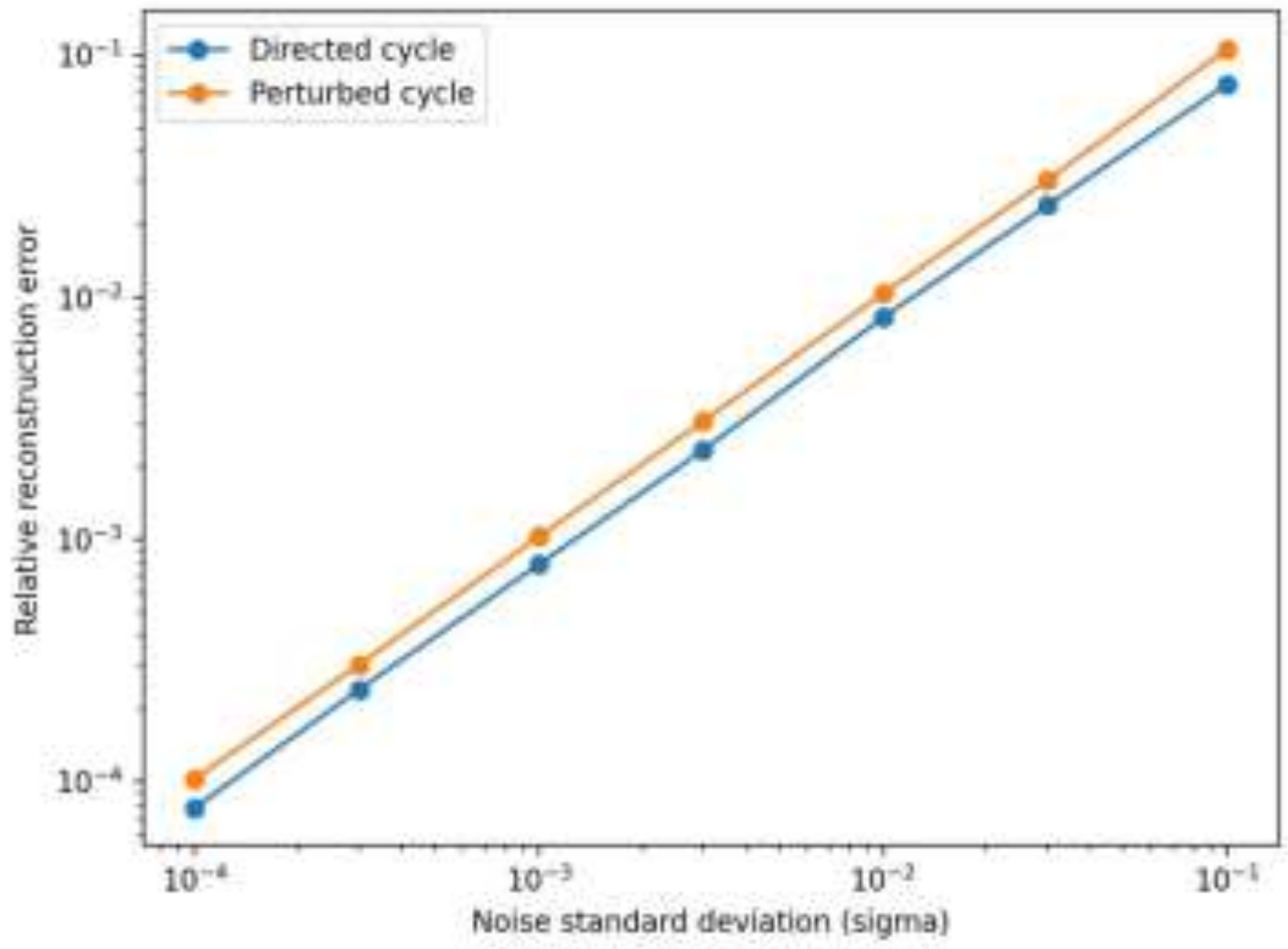

Figure 2: Reconstruction error versus input noise for normal (cycle) and non-normal (perturbed) digraphs. Stronger non-normality typically yields larger amplification in accordance with the conditioning factor $\kappa(V)$.

## 8. Conclusion

We developed a Laplacian-centric directed harmonic analysis based on the combinatorial directed Laplacian and a biorthogonal spectral calculus. The framework provides exact analysis/synthesis and diagonal filtering while explicitly quantifying metric distortion and instability mechanisms due to nonnormality. Directed variation defined by $\|Lx\|_2$ yields sharp BGFT-domain bounds, and sampling/reconstruction results separate sampling geometry from eigenvector geometry through explicit stability constants. Experiments confirm that filter and reconstruction robustness tracks eigenvector conditioning and departure-from-normality metrics, providing a principled "trust metric" for directed spectral methods.

**Acknowledgements**

The authors express their gratitude to the Commissioner of Collegiate Education (CCE), Government of Andhra Pradesh, and the Principal of Government College (Autonomous), Rajahmundry, for continued support and encouragement.

**Author Contributions**

The authors contributed equally in conceptualization, methodology, analysis, software, validation, and writing.

**Funding**

No external funding was received.

**Data Availability Statement**

No external datasets were used. The script that generates the figures (make_figures.py) is included for reproducibility; the author will also provide a public repository link upon acceptance.

**Conflicts of Interest**

The authors declares no conflicts of interest.